\documentclass[12pt]{article}
\pdfoutput =1
\usepackage{graphicx}
\DeclareGraphicsExtensions{.pdf}
\usepackage{float} 
\usepackage{amsmath}
\usepackage{amsfonts}   
\usepackage{amssymb}

\begin{document}
\date{}
\title{{\bf{\Large Analytic study of Gauss-Bonnet holographic superconductors in Born-Infeld electrodynamics}}}

\author{
{\bf {\normalsize Sunandan Gangopadhyay}$^{a,c,d}
$\thanks{sunandan.gangopadhyay@gmail.com, sunandan@bose.res.in}},
{\bf {\normalsize Dibakar Roychowdhury}
$^{b,}$\thanks{dibakar@bose.res.in}}\\
$^{a}$ {\normalsize Department of Physics, West Bengal State University, Barasat, India}\\
$^{b}$ {\normalsize  S.N. Bose National Centre for Basic Sciences,}\\{\normalsize JD Block, 
Sector III, Salt Lake, Kolkata 700098, India}\\[0.3cm]
$^{c}${\normalsize Visiting Associate in S.N. Bose National Centre for Basic Sciences,}\\
{\normalsize JD Block, Sector III, Salt Lake, Kolkata 700098, India}\\[0.3cm]
$^{d}${\normalsize Visiting Associate in Inter University Centre for Astronomy \& Astrophysics,}\\
{\normalsize Pune, India}\\[0.3cm]
}
\date{}

\maketitle

\begin{abstract}
{\noindent Using Sturm-Liouville (SL) eigenvalue problem, we investigate several properties of holographic $s$-wave superconductors in Gauss-Bonnet gravity with Born-Infeld electrodynamics in the probe limit. Our analytic scheme has been found to be in good agreement with the numerical results. From our analysis it is quite evident that the scalar hair formation at low temperatures is indeed affected by both the Gauss-Bonnet as well as the Born-Infeld coupling parameters. We also compute the critical exponent associated with the condensation near the critical temperature. The value of the critical exponent thus obtained indeed suggests a universal mean field behavior.}
\end{abstract}
\vskip 1cm

\section{Introduction}

The AdS/CFT correspondence \cite{adscft1}-\cite{adscft4} has been proved to be an extremely useful theoretical tool to understand strongly coupled field theories. In recent years, this holographic correspondence has been applied extensively to describe phase transitions in high $T_c$ superconductors. The model consists of a system with a AdS black hole and a charged scalar field minimally coupled to an abelian gauge field. The black hole admits scalar hair at a temperature $T$ below a certain critical temperature
$T_c$ by the mechanism of breaking of a local $U(1)$ symmetry near the event horizon of the black hole \cite{hs1}-\cite{hs5}. The emergence
of a hairy AdS black hole implies the formation of a charged scalar condensate in the dual CFTs by the AdS/CFT correspondence.

Till date a number of attempts have been made (mostly performed numerically) inorder to understand various properties of holographic superconductors in the
framework of usual Maxwell electromagnetic theory \cite{hs6}-\cite{hs24}. Recently, some investigations have also been performed in the framework of non-linear electrodynamics \cite{hs25}-\cite{hs22}. The first analytic computation of the effects of Born Infeld (BI) electrodynamics on $s$-wave superconductors has been put forward by us \cite{dibakar} based on the Sturm-Liouville (SL) eigenvalue problem. In this paper, we present analytical studies of Gauss-Bonnet holographic superconductors in BI electrodynamics upto first order in the Gauss-Bonnet and BI coupling parameter in the
probe limit. This means that we study the effect of the leading possible higher derivative corrections 
to the onset of $ s $-wave order parameter condensation. 

The physical meaning of looking at the leading order corrections of the BI coupling parameter is that we are
studying the effects due to higher derivative corrections of gauge fields on the $ s $-wave order parameter condensation. On the other hand, the study of the effects of the curvature corrections on $ (3+1) $-dimensional holographic superconductors is motivated by the application of the Mermin-Wagner theorem (which forbids continuous symmetry breaking in $ (2+1) $-dimensions because of large fluctuations in lower dimensions) to the holographic superconductors \cite{hs8}.
The physical meaning of looking at the leading order corrections of the Gauss-Bonnet coupling parameter is that we are
studying the effects due to higher derivative corrections of curvature on the $ s $-wave order parameter condensation.

Our results are in good agreement with the numerical computations carried out by us. We feel that this study is worthwhile since very little has been
done in problems involving non-linear electrodynamics both analytically as well as numerically.

Before going further let us now mention about the organization of our paper. In section 2 we give the basic setup for holographic superconductors in Gauss-Bonnet AdS gravity. In section 3 we explicitly compute the relationship between the critical temperature and charge density.  In section 4 we compute the condensation operator near the critical temperature. Finally we draw our conclusion in section 5.

\section{Basic set up for Gauss-Bonnet holographic superconductors in Born Infeld electrodynamics}
We perform our analysis in the fixed background of the Gauss-Bonnet AdS space time in the probe limit. We begin with the action of the Einstein Gauss-Bonnet gravity in five dimensions, which is given by \cite{hs11}
\begin{equation}
S=\int d^{5}x \sqrt{-g}(R+\frac{12}{l^{2}}+\frac{\alpha}{2}(R^{2}-4 R^{\mu\nu}R_{\mu\nu}+R^{\mu\nu\rho\sigma}R_{\mu\nu\rho\sigma}))\label{eq1}
\end{equation} 
where $ \frac{12}{l^{2}} $ stands for the cosmological constant and $ \alpha $ is the Gauss-Bonnet coefficient. At this stage it is to be noted that for the rest of our analysis we set $ l=1 $. The Ricci flat solution for the action (\ref{eq1}) is given by,
\begin{equation}
ds^{2}=-f(r)dt^{2}+f^{-1}(r)dr^{2}+r^{2}(dx^{2}+dy^{2}+dz^{2})
\end{equation} 
where,
\begin{equation}
f(r)=\frac{r^{2}}{2\alpha}\left(1-\sqrt{1-4\alpha\left(1-\frac{M}{r^{4}}\right) } \right). 
\end{equation}
Here $ M $ is the ADM mass of the black hole, which is related to the horizon radius ($ r_+ $) as,  $ r_+=M^{\frac{1}{4}} $. The Hawking temperature of the black hole is given by
\begin{equation}
T=\frac{r_+}{\pi}\label{temp}.
\end{equation} 

It is interesting to note that in the limit $ r\rightarrow\infty $ we have
\begin{equation}
f(r)\sim\frac{r^{2}}{2\alpha}\left[1-\sqrt{1-4\alpha} \right] 
\end{equation}
which naturally sets an effective radius for the AdS space time as,
\begin{equation}
L_{eff}^{2}=\frac{2\alpha}{1-\sqrt{1-4\alpha} }.
\end{equation}
The corresponding upper bound $ \alpha=\frac{1}{4} $ is known as Chern-Simons limit.

In order to study the $ s $ wave superconductors in the frame work of BI electrodynamics we adopt the following Lagrangian density which includes the Maxwell field ($ A_{\mu} $) and a charged complex scalar field ($ \psi $) as,
\begin{equation}
\mathcal{L}=\frac{1}{b}\left(1-\sqrt{1+\frac{b}{2}F^{\mu\nu}F_{\mu\nu}} \right)-|\nabla_{\mu}\psi- iA_{\mu}\psi |^{2}-m^{2}\psi^{2} 
\end{equation}
where $ b $ is the Born-Infeld parameter and $ m^{2} $ is the mass square of the scalar field.

With the following gauge choices for the vector field and the scalar field,  \cite{hs6}
\begin{eqnarray}
A_{\mu}=(\phi(r),0,0,0),~~~~~\psi=\psi(r)
\end{eqnarray}
we eventually arrive at the following equations of motion for the scalar potential $ \phi(r) $ and the scalar field $ \psi(r) $
\begin{equation}
\partial_{r}^{2}\phi + \frac{3}{r}(1-b(\partial_{r} \phi)^{2})\partial_{r}\phi -\frac{2\psi^{2}\phi}{f}(1-b(\partial_{r}\phi)^{2})^{3/2}=0\label{eq2}
\end{equation}
and,
\begin{equation}
\partial_{r}^{2}\psi + \left(\frac{3}{r}+\frac{\partial_{r}f}{f} \right)\partial_{r}\psi + \frac{\phi^{2}\psi}{f^{2}}-\frac{m^{2}\psi}{f}=0\label{eq3}. 
\end{equation}
In order to solve (\ref{eq2}) and (\ref{eq3}) one needs the boundary conditions near the event horizon $ r\sim r_+ $ and at the spatial infinity $ r\rightarrow\infty $. Considering the former case we have,
\begin{equation}
\phi(r_+)=0,~~~~~\psi(r_+)=\frac{\partial_{r}f(r_+)}{m^{2}}\partial_{r}\psi(r_+)
\end{equation}
whereas in the later case we have,
\begin{eqnarray}
\phi(r)&=&\mu - \frac{\rho}{r^{2}}\label{phi}\\
\psi(r)&=&\frac{\psi^{-}}{r^{\Delta_{-}}}+\frac{\psi^{+}}{r^{\Delta_{+}}}
\end{eqnarray}
where $ \mu $ and $ \rho $ are respectively the chemical potential and the charge density and $ \Delta_{\pm} =2\pm\sqrt{4+m^{2}L_{eff}^{2}}$ is the conformal dimension of the dual operator $ \mathcal{O} $ in the boundary field theory. In the following analysis we shall set $ \psi^{-}=0 $ and $ \psi^{+}=<\mathcal{O}> $. 


\section{Relation between critical temperature and charge density}
In order to obtain an explicit relationship between the critical temperature and the charge density we define $ z=\frac{r_+}{r} $. Under this choice of variables, we may express (\ref{eq2}) and (\ref{eq3}) as,
\begin{equation}
\partial_{z}^{2}\phi -\frac{1}{z}\partial_{z}\phi + \frac{3bz^{3}}{r_+^{2}}(\partial_{z}\phi)^{3}-\frac{2\psi^{2}\phi r_+^{2}}{f z^{4}}\left(1-\frac{b z^{4}(\partial_{z}\phi)^{2}}{r_+^{2}}\right)^{3/2}=0\label{eq4}
\end{equation}
and,
\begin{equation}
\partial_{z}^{2}\psi - \frac{1}{z}\partial_{z}\psi +\frac{\partial_{z}f}{f}\partial_{z}\psi + \frac{\phi^{2}\psi r_+^{2}}{z^{4}f^{2}} - \frac{m^{2}\psi r_+^{2}}{f z^{4}}=0\label{eq5}
\end{equation}
respectively.

At $ T=T_c $ , we have $ \psi=0 $. Therefore from (\ref{eq4}) we have
\begin{equation}
\partial_{z}^{2}\phi -\frac{1}{z}\partial_{z}\phi + \frac{3bz^{3}}{r_+^{2}}(\partial_{z}\phi)^{3}=0
\end{equation}
which has the solution,
\begin{equation}
\phi(z)=\lambda r_{+c}(1-z^{2})\left[1-\frac{b \lambda^{2}}{2}\xi(z) \right]+O(b^{2})\label{eq6} 
\end{equation}
with,
\begin{equation}
\xi(z)= (1+z^{2})(1+z^{4})~~~and,~~~\lambda = \frac{\rho}{r_{+c}^{3}}\label{lambda}.
\end{equation} 

 At this stage it is customary to mention that, since the BI coupling parameter ($ b $) is too small, therefore throughout the rest of our analysis we consider terms only linear in $ b $ and drop all the higher order terms in it.  

In order to investigate the boundary behavior of $ \psi $ (as $ T \rightarrow T_c $) we consider
\begin{equation}
\psi|_{z\rightarrow 0}\sim \frac{<\mathcal{O}>}{r_{+}^{3}}z^{3}F(z)\label{eq7}
\end{equation}
where $ F(0)=1 $ and $ F^{'}(0)=0 $.
Finally using (\ref{eq6}) and (\ref{eq7}), from (\ref{eq5}) we obtain
\begin{equation}
F^{''}+p(z)F^{'}+q(z)F+\lambda^{2}w(z)F=0\label{eq8}
\end{equation} 
where, the prime denotes the derivative w.r.t $ z $ and,
\begin{eqnarray}
p(z)&=&\frac{3(1-\sqrt{1-4\alpha + 4\alpha z^{4}})-12\alpha +20\alpha z^{4}}{z[1-4\alpha +4\alpha z^{4}-\sqrt{1-4\alpha + 4\alpha z^{4}}]}\\
q(z)&=&\frac{1}{z^{2}}\left[\frac{3(1-4\alpha -4\alpha z^{4}-\sqrt{1-4\alpha + 4\alpha z^{4}})}{\sqrt{1-4\alpha + 4\alpha z^{4}}-1+4\alpha -4\alpha z^{4}}+\frac{2m^{2}\alpha}{\sqrt{1-4\alpha + 4\alpha z^{4}}-1} \right] \\
w(z)&=&\frac{4\alpha^{2}(1-z^{2})^{2}(1-\frac{b}{2}\lambda^{2}\xi(z))^{2}}{(1-\sqrt{1-4\alpha + 4\alpha z^{4}})^{2}}.
\end{eqnarray}
It is now very much straightforward to convert (\ref{eq8}) into a standard Sturm-Liouville (SL) eigenvalue problem \cite{hs7} which reads,
\begin{equation}
(T(z)F(z))^{'}-Q(z)F(z)+\lambda^{2}P(z)F(z)=0
\end{equation} 
where,
\begin{eqnarray}
T(z)&=&\frac{z^{3}}{2\sqrt{\alpha}}(\sqrt{1-4\alpha + 4\alpha z^{4}}-1)\approx z^{3}\sqrt{\alpha}(z^{4}-1)[1-\alpha(z^{4}-1)]\\
Q(z)&=&-T(z)q(z)\approx -3z\sqrt{\alpha}(3z^{4}+6\alpha z^{4}-7\alpha z^{8})\\
P(z)&=&T(z)w(z)\approx \frac{\sqrt{\alpha}z^{3}(z^{2}-1)(1+\alpha(z^{4}-1))(1-\frac{b}{2}\lambda^{2}\xi(z))^{2}}{z^{2}+1}\label{eq9}
\end{eqnarray}
Here we have retained terms only upto an order $ \alpha^{3/2} $ while computing all the above expressions. It is quite interesting to note that one can further simplify (\ref{eq9}) using the fact that,
\begin{equation}
b\lambda^{2}=b(\lambda^{2}|_{b=0})+O(b^{2})
\end{equation} arXiv:1201.6520 [hep-th]
which finally yields,
\begin{equation}
P(z)\approx \frac{\sqrt{\alpha}z^{3}(z^{2}-1)(1+\alpha(z^{4}-1))(1-b(\lambda^{2}|_{b=0})\xi(z))}{z^{2}+1}.
\end{equation}

With all the above expressions in hand, the minimum value of $ \lambda^{2} $ may be obtained considering the variation of the following functional
\begin{equation}
\lambda^{2}[F(z)]=\frac{\int_{0}^{1}dz(T(z)(F^{'}(z))^{2}+Q(z)F^{2}(z))}{\int_{0}^{1}dz P(z) F^{2}(z)}
\end{equation}
with the choice $ F(z)=1-a z^{2} $ and $ m^{2}=-3/L_{eff}^{2} $. It is reassuring to note that this choice of mass is well above the BF bound \cite{bf1}-\cite{bf2}.

Finally using (\ref{temp}) and (\ref{lambda}) we obtain for $ T \sim T_c  $
\begin{equation}
T_c=\zeta \rho^{1/3}
\end{equation}
where $ \zeta=\frac{1}{\pi \lambda_{min}^{1/3}} $.

In the following, we tabulate various values of the coefficient ($ \zeta $) for different choice of parameters $ \alpha $  and $ b $. Looking at the results enumerated in tables (1 and 2) it is indeed evident that for $ b=0 $ case the results are in good agreement with the results existing in the literature \cite{hs8}, \cite{hs11}. Furthermore, we note that the condensation gets harder as the BI coupling parameter ($ b $) becomes larger. This result is consistent with the earlier findings \cite{hs22}. Finally as an effect of the Gauss-Bonnet coupling parameter ($ \alpha $) on the condensation value we note that, for a given value of $ b $, the condensation is harder to form for the larger value of $ \alpha $ \cite{hs8}, \cite{hs11},\cite{hs23}. In figure 1, we show the effect of Gauss Bonnet as well as BI coupling parameters on the critical temperature $ (T_c) $.  

\begin{figure}[h]
\centering
\includegraphics[angle=0,width=12cm,keepaspectratio]{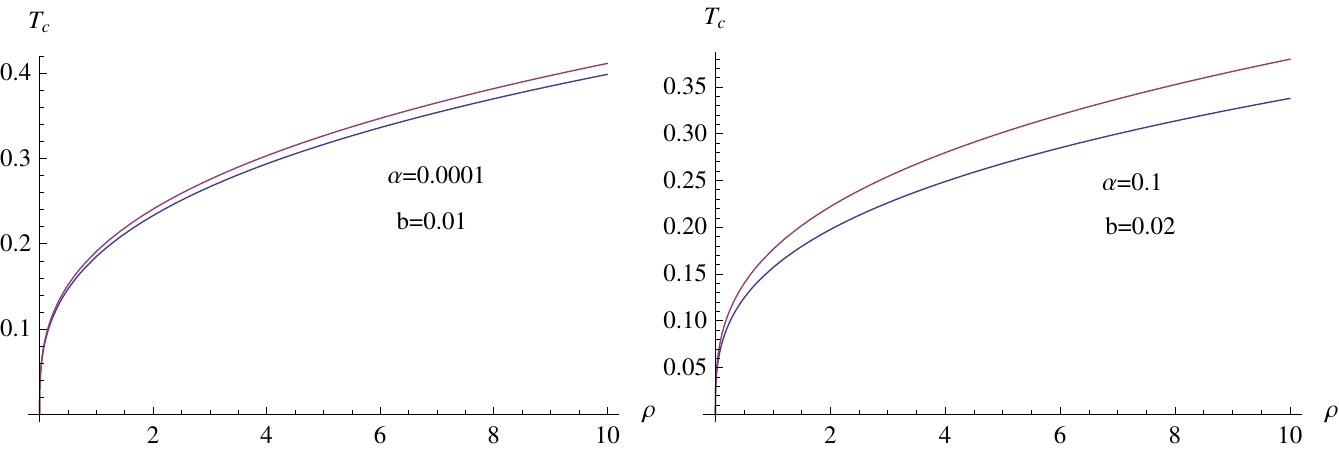}
\caption[]{\it Critical temperature ($ T_{c}-\rho $) plot for Gauss Bonnet holographic superconductors for different choice of parameters $ \alpha $ and $ b$  $ $. The red (upper) curve corresponds to the numerical value whereas the blue (lower) one corresponds to analytical value. From the plots it is evident that the critical temperature is indeed suppressed due to the higher curvature effects.}
\label{figure 2a}
\end{figure}

\begin{table}[htb]
\caption{A comparison of the analytical and numerical results for the critical temperature for $ \alpha=0.0001 $}   
\centering                          
\begin{tabular}{c c c c c c c}            
\hline\hline                        
$b$ & $ a $& $ \lambda_{min}^{2} $ & $\zeta_{SL}\left(=\frac{1}{\pi \lambda_{min}^{1/3}}\right)$  & $\zeta_{Numerical}$ &  \\ [0.05ex]
\hline
0 & 0.721772 & 18.2331 & 0.196204& 0.196204 \\
0.01 &0.754014 & 25.9147 & 0.185037&0.191012  \\
0.02 & 0.821082 &44.0958& 0.169349 &0.185073  \\ [0.5ex] 
\hline                              
\end{tabular}\label{E1}  
\end{table}
\begin{table}[htb]
\caption{A comparison of the analytical and numerical results for the critical temperature for $ \alpha=0.1 $}   
\centering                          
\begin{tabular}{c c c c c c c}            
\hline\hline                        
$b$ & $ a $& $ \lambda_{min}^{2} $ & $\zeta_{SL}\left(=\frac{1}{\pi \lambda_{min}^{1/3}}\right)$  & $\zeta_{Numerical}$ &  \\ [0.05ex]
\hline
0 & 0.709061 & 21.5679 & 0.19078& 0.189939 \\
0.01 &0.754297 & 33.3478 &0.177421&0.184679  \\
0.02 &0.868304 &70.2223& 0.156713 &0.17616  \\ [0.5ex] 
\hline                              
\end{tabular}\label{E1}  
\end{table}


\section{Critical exponent and condensation values}

In this section we aim to examine the effect of BI coupling parameter on the condensation operator near the critical point.
In order to do that we first note that close to the critical temperature ($ T_c $) equation (\ref{eq4}) may be written as,
\begin{equation}
\partial_{z}^{2}\phi -\frac{1}{z}\partial_{z}\phi + \frac{3bz^{3}}{r_+^{2}}(\partial_{z}\phi)^{3}=\frac{<\mathcal{O}>^{2}}{r_+^{4}}\mathcal{B}\phi \label{eq10}
\end{equation}
where, $ \mathcal{B}=\frac{2z^{2}}{f}\left(1-\frac{b z^{4}(\partial_{z}\phi)^{2}}{r_+^{2}}\right)^{3/2}F^{2}(z) $.

Since $ \frac{<\mathcal{O}>^{2}}{r_+^{4}} $ is a very small parameter (as we are very close to the critical temperature), therefore it will be natural to expand $ \phi(z) $ as,

\begin{equation}
\frac{\phi(z)}{r_+}=\lambda (1-z^{2})\left[1-\frac{b \lambda^{2}}{2}\xi(z) \right]+ \frac{<\mathcal{O}>^{2}}{r_+^{4}} \chi(z)\label{eq11}
\end{equation}
with $ \chi(1)=\chi^{'}(1)=0 $.

Using (\ref{eq11}), from (\ref{eq10}) we obtain,
\begin{equation}
\chi^{''}(z)-\frac{\chi^{'}}{z}+36b\lambda^{2}z^{5}\chi^{'}=\lambda \mathcal{B}(1-z^{2})\left( 1-\frac{b}{2}\lambda^{2}\xi(z)\right)\label{eq12}. 
\end{equation}
From (\ref{eq12}) we note that, in the limit  $ z\rightarrow0 $
\begin{equation}
\chi^{''}(0)=\frac{\chi^{'}(z)}{z}|_{z\rightarrow 0}\label{eq13}.
\end{equation} 

In order to obtain the r.h.s. of (\ref{eq13}), we note that equation (\ref{eq12}) may be written as,
\begin{equation}
\frac{d}{dz}\left(e^{6b\lambda^{2}z^{6}}\frac{\chi^{'}}{z} \right)=\lambda \frac{2z^{3}}{r_+^{2}}\frac{e^{6b\lambda^{2}z^{6}}(1-\frac{b}{2}\lambda^{2}\Gamma(z))}{(1+z^{2})(1+\alpha(1-z^{4}))}F^{2}(z)\label{eq14} 
\end{equation}
where, $ \Gamma(z)=1+z^{2}+z^{4}+13z^{6} $.
Finally integrating (\ref{eq14})  between the limits $ z=0 $ and $ z=1 $ and keeping terms only linear in $ b $, we obtain
\begin{eqnarray}
\frac{\chi^{'}}{z}|_{z\rightarrow 0}&=&-\frac{\lambda}{r_+^{2}}\mathcal{A}\label{eq15}\\
with,~~~~\mathcal{A}&\approx &\int_{0}^{1}\frac{2z^{3}F^{2}(z)(1-\frac{b}{2}\lambda^{2}(1+z^{2}+z^{4}+z^{6}))(1-\alpha(1-z^{4}))}{1+z^{2}}dz\nonumber.
\end{eqnarray}
 
Again from (\ref{phi}) and (\ref{eq11}) one can have near $ z=0 $
\begin{equation}
\mu - \frac{\rho}{r_+^{2}}z^{2}=\lambda r_{+}(1-z^{2})\left[1-\frac{b \lambda^{2}}{2}\xi(z) \right]+\frac{<\mathcal{O}>^{2}}{r_+^{3}}(\chi(0)+z\chi^{'}(0)+\frac{z^{2}}{2}\chi^{''}(0)+...).
\end{equation}
Comparing the coefficients of $ z^{2} $ from both sides and using (\ref{temp}), (\ref{lambda}), (\ref{eq13}) and (\ref{eq15}) for $ T \rightarrow T_c $ we finally obtain,
\begin{eqnarray}
<\mathcal{O}> &\approx & \gamma \pi^{3}T_c^{3}\sqrt{1-\frac{T}{T_c}}\\
with,~~~~\gamma &=& \sqrt{\frac{6}{\mathcal{A}}}.
\end{eqnarray}
 
Let us now tabulate various values of the coefficient $ \gamma $ those obtained numerically as well as analytically using SL eigenvalue problem.

\begin{table}[htb]
\caption{A comparison of the analytical and numerical results for the condensation operator for $ \alpha=0.0001 $}   
\centering                          
\begin{tabular}{c c c c c c c}            
\hline\hline                        
$b$ & $ a $& $ \lambda_{min}^{2} $ & $\gamma_{SL}$  & $\gamma_{Numerical}$ &  \\ [0.05ex]
\hline
0 & 0.721772 & 18.2331 & 7.70525& 7.70677 \\
0.01 &0.754014 & 25.9147 & 8.73543&8.7599  \\
0.02 & 0.821082 &44.0958& 10.36 &11.1853  \\ [0.5ex] 
\hline                              
\end{tabular}\label{E1}  
\end{table}

\begin{table}[htb]
\caption{A comparison of the analytical and numerical results for the condensation operator for $ \alpha=0.1 $}   
\centering                          
\begin{tabular}{c c c c c c c}            
\hline\hline                        
$b$ & $ a $& $ \lambda_{min}^{2} $ & $\gamma_{SL}$  & $\gamma_{Numerical}$ &  \\ [0.05ex]
\hline
0 & 0.709061 & 21.5679 & 7.90303& 7.86604 \\
0.01 &0.754297 & 33.3478 &9.26635&9.3564  \\
0.02 &0.868304 &70.2223& 11.7452 &14.4142 \\ [0.5ex] 
\hline                              
\end{tabular}\label{E1}  
\end{table}


\newpage
\section{Conclusions}
In this paper, we have made an analysis on holographic $s$-wave superconductors in the 
background of Gauss-Bonnet (GB) AdS spacetime. Following our methodology in \cite{dibakar}, we perform our analysis both analytically as well as numerically. Using Sturm-Liouville (SL) eigenvalue problem we establish the relationship between the critical temperature and the charge density. We show that both the values of BI coupling parameter ($b$) as well GB coupling parameter ($\alpha$) (at the leading order) indeed affects the formation of scalar hair at low temperatures. The physical meaning of the leading order expansions in $b$ and $\alpha$ is that what is being studied is the effect of leading higher derivative corrections of gauge fields and curvature to the onset of the $s$-wave order parameter condensation.
It is observed that the condensation is harder to form for higher values of BI/GB coupling parameters. The critical exponent associated with the condensation again comes out to be $ 1/2 $ which is in good agreement with the universal mean field value.

\section*{Acknowledgments} DR would like to thank CSIR for financial support. SG would like to thank his wife Dr.(Mrs) Shreemoyee Ganguly for useful
discussions and extending her generous help with computational problems.



\begin{thebibliography}{99}
\baselineskip=0.6 cm
\bibitem{adscft1} J. M. Maldacena,  Adv. Theor. Math. Phys. 2, 231 (1998).[hep-th/9711200].
\bibitem{adscft2} E. Witten,  Adv. Theor. Math. Phys. 2, 253 (1998).
\bibitem{adscft3} S. S. Gubser, I. R. Klebanov, and A. M. Polyakov,  Phys. Lett. B 428, 105 (1998).
\bibitem{adscft4} O. Aharony, S. S. Gubser, J. M. Maldacena, H. Ooguri, and Y. Oz,  Phys. Rept. 323, 183 (2000).
\bibitem{hs1} S. S. Gubser, Class. Quant. Grav.  22, 5121 (2005).
\bibitem{hs2} S. S. Gubser,  Phys. Rev. D 78, 065034 (2008).
\bibitem{hs3} S.A. Hartnoll, Class. Quantum Grav. 26, 224002 (2009).
\bibitem{hs4} C.P. Herzog,  J. Phys. A  42, 343001 (2009).
\bibitem{hs5} G. T. Horowitz,  arXiv:1002.1722 [hep-th] (2010).
\bibitem{hs6} S. A. Hartnoll, C. P. Herzog, G. T. Horowitz,  Phys. Rev. Lett. 101, 031601 (2008).
\bibitem{hs7} G. Siopsis, J. Therrien, JHEP 05, 013 (2010).
\bibitem{hs8} R. Gregory, S. Kanno, J. Soda,  JHEP 0910 (2009) 010.
\bibitem{hs9} S. A. Hartnoll, C. P. Herzog, G. T. Horowitz, JHEP  0812, 015 (2008).
\bibitem{hs10} H. B. Zeng, X. Gao, Y. Jiang, H. S. Zong, JHEP 1105 (2011) 002.
\bibitem{hs11} H. F. Li, R. G. Cai, H. Q. Zhang, JHEP 1104 (2011) 028.
\bibitem{hs12} R. G. Cai, H. F. Li, H. Q. Zhang, Phys. Rev. D 83, 126007 (2011).
\bibitem{hs14}  Q. Y. Pan, B. Wang, E. Papantonopoulos, J. Oliveira, A. Pavan, Phys. Rev. D 81,106007 (2010).
\bibitem{hs15}  R.G. Cai, H. Zhang, Phys. Rev. D 81, 066003 (2010).
\bibitem{hs16}G. T. Horowitz, M. M. Roberts, Phys. Rev. D 78, 126008 (2008).
\bibitem{hs17} G. T. Horowitz, M. M. Roberts, JHEP  0911 (2009) 015.
\bibitem{hs18}  Q Pan, J. Jing, B. Wang, JHEP 1111 (2011) 088.
\bibitem{hs23} R. G. Cai, Z. Y. Nie, H. Q. Zhang, Phys. Rev. D 82, 066007 (2010).
\bibitem{hs24} Q. Pan, B. Wang, Phys. Lett. B 693 (2010) 159.
\bibitem{hs25} Y. Liu, Y. Peng, B. Wang, arXiv:1202.3586 [hep-th].
\bibitem{hs19} J. Jing, S. Chen, Phys. Lett. B 686 (2010) 68.
\bibitem{hs20} J. Jing, Q Pan, S. Chen, JHEP 1111 (2011) 045. 
\bibitem{hs21} J. Jing, Q Pan,  B. Wang, Phys. Rev. D 84,126020, (2011).
\bibitem{hs22} J. Jing,  L. Wang, Q Pan, S. Chen, Phys.Rev.D 83,066010, (2011).
\bibitem{dibakar} S. Gangopadhyay, D. Roychowdhury, arXiv:1201.6520 [hep-th], (To appear in JHEP).
\bibitem{bf1}P. Breitenlohner, D. Z. Freedman, Phys. Lett. 115B, (1982) 197.
\bibitem{bf2}P. Breitenlohner, D. Z. Freedman, Ann. Phys. 144 (1982) 197.
\end{thebibliography}
\end{document}